\documentclass[sigconf,natbib=true,anonymous=false]{acmart}

\usepackage{amsmath,amsfonts}
\usepackage{algorithmic}
\usepackage{graphicx}
\usepackage{textcomp}
\usepackage{xcolor}

\usepackage{amsmath,amsfonts}
\usepackage{physics}
\usepackage{bm}

\usepackage{array}
\usepackage{multirow}
\usepackage{tabularx}
\usepackage{graphicx}
\usepackage{caption3}
\usepackage{color, colortbl}
\usepackage{tcolorbox}
\usepackage{siunitx}
\usepackage{subcaption}
\usepackage{listings}
\usepackage{float}
\usepackage{makecell}
\usepackage{tabularx}
\usepackage[flushleft]{threeparttable}

\AtBeginDocument{%
  \providecommand\BibTeX{{%
    \normalfont B\kern-0.5em{\scshape i\kern-0.25em b}\kern-0.8em\TeX}}}

\copyrightyear{2023} 
\acmYear{2023} 
\setcopyright{acmlicensed}\acmConference[SIGIR '23]{Proceedings of the 46th International ACM SIGIR Conference on Research and Development in Information Retrieval}{July 23--27, 2023}{Taipei, Taiwan}
\acmBooktitle{Proceedings of the 46th International ACM SIGIR Conference on Research and Development in Information Retrieval (SIGIR '23), July 23--27, 2023, Taipei, Taiwan}
\acmPrice{15.00}
\acmDOI{10.1145/3539618.3591831}
\acmISBN{978-1-4503-9408-6/23/07}

\begin{document}

\title{Integrity and Junkiness Failure Handling for Embedding-based Retrieval: A Case Study in Social Network Search}


\author{Wenping Wang}
\affiliation{%
  \institution{Meta Platforms, Inc}
  \streetaddress{1 Hacker Way}
  \city{Menlo Park}
  \state{California}
  \country{USA}}
\email{wenping@meta.com}

\author{Yunxi Guo}
\affiliation{%
  \institution{Meta Platforms, Inc}
  \streetaddress{1 Hacker Way}
  \city{Menlo Park}
  \state{California}
  \country{USA}}
\email{yguo0418@meta.com}

\author{Chiyao Shen}
\affiliation{%
  \institution{Meta Platforms, Inc}
  \streetaddress{1 Hacker Way}
  \city{Menlo Park}
  \state{California}
  \country{USA}}
\email{chiyao@meta.com}

\author{Shuai Ding}
\affiliation{%
  \institution{Meta Platforms, Inc}
  \streetaddress{1 Hacker Way}
  \city{Menlo Park}
  \state{California}
  \country{USA}}
\email{sding@meta.com}

\author{Guangdeng Liao}
\affiliation{%
  \institution{Meta Platforms, Inc}
  \streetaddress{1 Hacker Way}
  \city{Menlo Park}
  \state{California}
  \country{USA}}
\email{guangdeng@meta.com}

\author{Hao Fu}
\affiliation{%
  \institution{Meta Platforms, Inc}
  \streetaddress{1 Hacker Way}
  \city{Menlo Park}
  \state{California}
  \country{USA}}
\email{haofu@meta.com}

\author{Pramodh Karanth Prabhakar}
\affiliation{%
  \institution{Meta Platforms, Inc}
  \streetaddress{1 Hacker Way}
  \city{Menlo Park}
  \state{California}
  \country{USA}}
\email{pramodh@meta.com}


\begin{abstract}

Embedding based retrieval has seen its usage in a variety of search applications like e-commerce, social networking search etc. While the approach has demonstrated its efficacy in tasks like semantic matching and contextual search, it is plagued by the problem of uncontrollable relevance. In this paper, we conduct an analysis of embedding-based retrieval launched in early 2021 on our social network search engine, and define two main categories of failures introduced by it, integrity and junkiness. The former refers to issues such as hate speech and offensive content that can severely harm user experience, while the latter includes irrelevant results like fuzzy text matching or language mismatches. Efficient methods during model inference are further proposed to resolve the issue, including indexing treatments and targeted user cohort treatments, etc. Though being simple, we show the methods have good offline NDCG and online A/B tests metrics gain in practice. We analyze the reasons for the improvements, pointing out that our methods are only preliminary attempts to this important but challenging problem. We put forward potential future directions to explore.

\end{abstract}




\maketitle

\section{Introduction}
Traditional search engines rely a lot on lexical matching for retrieving relevant results from a given query (mostly reverted index for documents). However this approach has the well known shortcoming in terms of semantic match and personalized matching, which are common and vital in the success of commercial search products like amazon, google, etc. Starting from facebook’s approach~\cite{huang2020embedding}, embedding based retrieval is developed to deal with these scenarios, and have proved their success in real world business like amazon~\cite{nigam2019semantic}, taobao ~\cite{li2021embedding}, facebook~\cite{huang2020embedding}, jingdong~\cite{zhang2020towards}, baidu~\cite{liu2021pre} etc. Though having different assumptions, improvements in the deep model training and use cases, these embedding systems all share a similar two tower structure. During training, the two tower model is trained for user and candidate together, with various use case specific embeddings like user context, query n-gram, etc as inputs. A similarity score is computed as distance between two embeddings. In online inference, candidate embedding is pre-calculated and saved into index, while user side embedding is computed on the fly.


Such approaches, though having proved their success in industry, suffer from low controllability of relevance. From TaoBao’s learning, the “inability to exactly matching query terms”~\cite{li2021embedding} has caused user complaints and bad cases in production. Based on the survey result on our social network search product, more than half of the negative user feedback are caused by embedding-based retrieval. The issue greatly hurts the user experience. Meanwhile, it is not an easy task to address these failures. Few previous works, with their success stories, have explicitly called out the issue. Taobao~\cite{li2021embedding}, put forward a relevance control module which applies another “and” with the retrieved results. The method drops the unmatched results with query, requiring results to have both embedding similarity and text match, greatly limiting the embedding retrieval’s power in finding semantic matching or personalized results for users.

In this paper, we start from the status of failures on the platform generated by embedding-based retrieval (EBR), classifying the failures into two main categories, integrity and junkiness. The former refers to hard failures of embedding-based retrieval like misinformation, hate speech, porny results while the latter refers to general irrelevance between query and doc. We then propose separate treatments that are applied during model inference. For integrity violation cases, we apply strict clean up in index building. For junkiness feedback, we purpose customized discarding threshold method that discarding EBR retrieved results in a customized way, and triggering condition control of EBR. Being simple and generic to the two tower structure, we show with both offline metric and online A/B test results that it gives us surprisingly good metric wins. In the end, we point out there is still a long way to go in the domain and several potential future directions. The purposed methods have been applied in our search engine production since 2022.

The work is based on our social network search engine product - group search, where we have a baseline launch of the embedding based retrieval in early 2021 based on ~\cite{huang2020embedding}. The group refers to a product on our social network platform, providing a community of people with similar interests, goals or a shared target a channel to engage and explore. The search engine has hundreds of millions of queries everyday, and one of the main use cases are finding directly connected group and exploring unconnected groups. For downstream actions, \emph{Connected Navigation}, denoted as \textbf{CN}, refers to user being \textbf{navigate} to targeted connected group after clicking, while \emph{Unconnected Navigation}, denoted as \textbf{UN}, represents for users \textbf{join} a group.

Our goal of the paper is mainly to draw attention to the problem of EBR generated failures in real world search applications, in the setting of social network search especially, and provide our preliminary attempts to the problem. The rest of the paper is organized as follows. In the next section, we provide an overview of the initial status of the problem on our search engine. Section 3 talks about our methodology in detail. Section 4 discusses the metrics we used for offline evaluation, and the online A/B test setting with the results. We discuss the remaining problem in the end of Section 4. We conclude our work in the last section and point out the potential future directions for improvement.

\section{EBR status analysis}
We use search log to generate queries and ask human raters to rate our search engine returned sessions of results. (Section~\ref{sec:expts} gives detailed offline metrics for analysis). We selected 15k queries, representing major query intents and languages.

\subsection{Baseline Model Architecture}\label{sec:baseline}
Our baseline model implemented and launched in 2021 is based on the system in ~\cite{huang2020embedding}, and detailed model architecture in ~\cite{liu2021que2search}. We adopted several changes in input compared with ~\cite{liu2021que2search} that resulted in good online performance. 

For query tower, we added country, region, query text and trigram. The first three embedding dim is 16 and the latter two are 128. For doc tower, doc title, description, title trigram, language, region, topic, country are used. The title, description and trigram has dim 128 while the rest has dim 32. Query text, doc title, doc description have the top importance in our model. We used batch negative sampling and scaled loss (a multiplier on model output  in loss function) during training. We gather training data from search log. The rest details remains the same as ~\cite{liu2021que2search}. Our model AUC is 0.79. The baseline launch impact is shown in the bottom of table~\ref{tab:online_result}.

\subsection{Results}
\subsubsection{Total failure rate and EBR trigger rate}
The total prod failure rate are 7.3\%, 32.6\%, and 43.1\% at top 1, 3, and 5. By checking the source of the generated failure results, roughly 70\% are generated by the EBR node during the retrieval stage, for our group search product.

\subsubsection{Break down failure reason}
Table~\ref{tab:failures} shows the reasons of failure and corresponding percentage.
\begin{table}
  \caption{Distribution of EBR failures}
  \label{tab:failures}
  \begin{tabular}{ccl}
    \toprule
    Failure reasons&Percentage&Category\\
    \midrule
    irrelevant result (fuzzy text match) & 53\%& junkiness\\
    Location mismatch & 18\%& junkiness\\
    Language mismatch & 4\%& junkiness\\
    Misinformation & 10\%& integrity\\
    Untrustworthy results & 10\%& integrity\\
    Offensive results & 5\%& integrity\\
  \bottomrule
\end{tabular}
\end{table}

\subsection{Failure Categories}

From the practice of our selected social network search system, we define the failures into two levels,

\begin{itemize}
\item \textbf{junkiness failures}: irrelevant results given clear query intent, like random people returned given exact match connected friend name query
\item \textbf{integrity failures}: serious failures that would hurt users severely, like porny result, hate speech, fake news, etc
\end{itemize}

The two main categories of Embedding-based retrieval failures differs mostly in the severity. Ideally we would want search engine to provide as few irrelevant results as possible, and no integrity violations at all. In practice, Embedding-based retrieval serves those semantic matching case while generating a lot of irrelevant results at the same time. 

\section{methodology}

In the below sections, we purpose several treatments  after the two tower model (in Section~\ref{sec:baseline}) is trained. The baseline model uses cosine similarity to compute the similarity score and apply a threshold to either keep or reject the retrieved results.

\subsection{Junkiness Failure Reduction}

\subsubsection{Sigmoid}\label{sec:sigmoid}
One way to control the relevance of output is to view the similarity score as a probability measure of document relevance for a given user + query (user and query embeddings are in the query tower), i.e. 
\begin{displaymath}
  s\_score = F(Q, D) = P(relevant|user+query, doc)
\end{displaymath}

This naturally remind us of the sigmoid function, as its basic probabilistic intuition are similar. The idea behind is though we discard results if they fall below a threshold of similarity, different (user+query, doc) pair have different best trade-off point, for selected query intent, user cohort and doc topic. Applying sigmoid help centering the output distribution across different input segments, and transform those high confidence predictions, either positive or negative, towards 1 or 0.

We choose the logistic function for its simplicity in implementation

\begin{displaymath}
transformed\_score=g(s\_score) = \frac{1}{1 + e^{-s\_score}}
\end{displaymath}

the \emph{s\_score} refers to the output from similarity function, and depending on the specific model and application case, we apply different linear transform before applying the sigmoid function.

\subsubsection{Discarding threshold customization}
We can go further from the idea in 3.1.1. Instead of centering the output distribution, we seek to discard retrieved results in a customized way for different segments of inputs, i.e. a customized threshold for different user+query segment. This is actually the same topic in the domain of machine learning fairness ~\cite{Holstein2019}~\cite{du2020fairness}~\cite{li2021user}.

The intuition behind is for application like social network search, the potential candidate size number far exceeds the allowed number of results to return on mobile or web. Therefore the goal of the EBR design is to pick top semantic matching items in a very big candidate pool, i.e. precision driven. In our practice, we fit a simple linear regression model to do the customization.

\begin{displaymath}
Threshold=\sum_{j=0}^{n}{\boldsymbol\beta_j \cdot \mathbf{x}_{ij}}=\mathbf{y}_i
\end{displaymath}

${x}_{ij}$ above represents the input feature j for result i we believe our two tower model behaves differently on. We use user country, language, query intent and doc source type (CN or UN) to fit as input. ${y}_{i}$ is a threshold that discards results confidently. We use the following method to choose it. The trained model is first launched into production for a short period without any discarding (i.e. threshold being -inf) to gather logging data. With these data, we select from those results that have click or downstream action (group join or post comment after join, etc), and apply a threshold ${y}_{ip}$ such that top p\% of these results are left. For example, all the clicked unconnected English groups in US with people name intent query have similarity score within [0.2, 1.0], while the ${y}_{i30}$ threshold is 0.7.

The hyper parameter p is chosen globally via A/B test. In our practice, p is 0.9. We use MSE (mean squared error) to compute the \textbf{$\beta$}. We will show in section~\ref{sec:expts} that this method contributes most to the online evaluation metric win.

\subsubsection{Triggering control}
Another observation is that EBR does not fit for some clear query intents, for instance, directly connected celebrity finding intents. Mismatch result types also negates the usage, like friend photo intent with group EBR triggered. Such hypothesis can be easily verified by the ${y}_{ip}$ distribution on the given scenarios.

We disable the EBR sources completely for such cases found in analysis, and delegate the text based retrieval to handle the search request in replace. Examples are in section~\ref{sec:expts}.

\subsection{Integrity Failure Reduction}

Unlike junky results, integrity violation results are more critical to a user-facing search engine. In general we should avoid such cases from happening at all, even at some cost. Once a result is tagged as integrity sensitive, a ground truth label is saved in database. With this label, we apply

\begin{itemize}
\item \textbf{ranking filtering}: for edge cases lies between recommendable and non-recommendable. Examples like hate speech on opinion seeking topics, etc. We apply rule based demotion or model based demotion in ranking layers for this cases.
\item \textbf{indexing removal}: results that should not be returned at any time. Examples like child pornography, etc. We directly remove these pre-computed embedding from index to avoid retrieving at all.
\end{itemize}

\begin{table*}[!th]
    \centering
\linespread{1.1}\selectfont
\begin{center}
\centering
\(
    \begin{tabular}{l|rrrr}\hline
       Method  &$\Delta$ NDCG@1 &$\Delta$ NDCG@3 &$\Delta$ NDCG@5 &$\Delta$ NONREC \\\hline
       Sigmoid on Unconnected Navigation    &\textbf{3.523\%}   &{1.835\%}  &\textbf{2.232\%} &{0.26\%} \\\hline
       Threshold Customization - Connected Navigation &{0.261\%} &{0.653\%}   &\textbf{0.914\%} &\textbf{-1.04\%} \\\hline
       Threshold Customization - Unconnected Navigation &{0.01\%} &\textbf{10.591\%}   &\textbf{16.232\%} &{-0.88\%} \\\hline
       Triggering Control - Unconnected Navigation (person name query)    &\textbf{6.575\%} &\textbf{6.343\%}   &\textbf{5.631\%}  &{-0.65\%} \\\hline
       Triggering Control - Unconnected Navigation (US query only)   &\textbf{6.849\%} &\textbf{5.951\%}   &\textbf{5.243\%} &{-0.23\%} \\\hline
       Index Removal    &{-1.654\%}   &{-0.091\%}   &{0.394\%} &\textbf{-10.80\%} \\\hline
    \end{tabular}
    \)
    \caption{\label{tab:ndcg} The offline ndcg metrics comparison between Control and Test for each treatment. The number in boldface means that it is statistically significant ($p<0.05$) compared to the Control.}
    \end{center}
\end{table*}

\begin{table*}[!th]
    \centering
\linespread{1.1}\selectfont
\begin{center}
\centering
\(
    \begin{tabular}{l|rrr}\hline
       Method  &$\Delta$ {CLK} &$\Delta$ {CTR} &$\Delta$ {GJ}      \\\hline
       Sigmoid on Unconnected Navigation    &\textbf{0.55\%}   &\textbf{0.60\%}  &\textbf{0.12\%}       \\\hline
       Threshold Customization - Connected Navigation &{-0.03\%} &\textbf{0.86\%}   &\textbf{0.92\%}        \\\hline
       Threshold Customization - Unconnected Navigation &\textbf{0.30\%} &\textbf{0.34\%}   &{0.06\%}        \\\hline
       Triggering Control - Unconnected Navigation (person name query)   &{-0.13\%} &{0.02\%}   &\textbf{-0.27\%}        \\\hline
       Triggering Control - Unconnected Navigation (US query only)   &\textbf{-0.06\%} &\textbf{-0.02\%}   &\textbf{-0.10\%}        \\\hline
       Baseline   &\textbf{1.31\%} &\textbf{-1.06\%}   &\textbf{0.76\%}        \\\hline
    \end{tabular}
    \)
    \caption{\label{tab:online_result} The online engagement metrics comparison between Control and Test for each treatment. The number in boldface means that it is statistically significant ($p<0.05$) compared to the Control.}
    \end{center}
\end{table*}

\section{Experiments and Analysis}\label{sec:expts}
In this section, we report the statistics and analysis of offline and online metrics for our methods. We show that our treatment have surprisingly good results in failure reduction and online metric win, and the scale of online metric win is comparable to that of the provided baseline.

\subsection{Evaluation Metric}
\subsubsection{Offline metrics for failure control}
Relevance metric are widely used in search engines as offline metrics for quick development purpose. For junkiness failures, we adopted Normalized Discounted Cumulative Gain (NDCG) as offline evaluation measure to give the relevance labels for top 5 ranked (query, result) pairs. Results will be calculated on a user session level, i.e. given user and query, whether a session of returned result is better in relevance, denoted as \textbf{NDCG}. For integrity failures, the top 10 results will be checked in a user session and mark it to be integrity violated if one of the results fails, denoted as \textbf{NONREC} for being not recommendable.

\subsubsection{Online Metrics as Golden Targets}

The most important metrics for social network product are the online metrics, which are directly related to app's popularity and revenue. When experiment goes to the online A/B test, we use online metrics to judge whether a social network app is really improved by the experiment. We selected below representative online metrics in our experiments:
\begin{enumerate}
\item ~\textbf{Click and Click though rate}: Click measures total result click count while Click though rate measures the ratio of users clicks vs search results shown counts. Improving the click through rate can make the search result page to be more effective. Click is denoted as \textbf{CLK}. Click though rate is denoted as \textbf{CTR}.
\item ~\textbf{Search referred connections}: this metric measures how many connections are built via search engine. Users can add friends/join groups/following pages via people/group/page search, respectively. An improved search results can bring more connections for social network. This metric can be statistically significantly moved via a single experiment in several days. Denoted as \textbf{GJ} for the group join case.
\end{enumerate}

\subsection{Results and Findings}\label{sec:traffic_split}
We conduct the experiments during second half of 2022 from July to Dec. Each experiment will allocate 4\% of total user traffic and split to control and test groups randomly. For single experiment, we monitor the offline and online metrics for at least one week and make decision whether to launch it based on overall metric evaluations. We read and compare the metrics between Test and Control user groups to represent the performance of each launch.

\subsubsection{Offline experiment results}
In Table~\ref{tab:ndcg}, we show the comparison results with the above methods. The downward trend of junkiness reduction is clear.


\subsubsection{Online A/B test}
In Table~\ref{tab:online_result}, we show the results of different methods' online metric impact. We add a row in the bottom in the table~\ref{tab:online_result} to provide the online metric movement during the baseline launch of our group search EBR in early 2021.



Two key insights can be drawn from the above results.

\begin{enumerate} 
\item Failure handling in EBR is an important issue in terms of both offline and online metric.
\item Though majority of them target on discarding results, our purposed methods are effective in practice.
\end{enumerate}

The first insight is from comparing online metrics in the last row and the rest rows in Table~\ref{tab:online_result}, we can see the accumulated online impact of failure handling methods are comparable to the baseline. Meanwhile, according to Table~\ref{tab:ndcg} row 4 and Table~\ref{tab:online_result} row 4, with the NDCG improvement on Unconnected Navigation, there is a statistically significant improvement on online metrics.

\subsection{Jobs To Be Done}
Given the positive results of our purposed solutions, the journey is not yet finished. We compare the user survey conducted between the end of 2021 and the end of 2022 (during which we launched the solutions in the paper), the junkiness and integrity failures introduced by EBR remain to be a big concern. For instance, the location mismatch dropped by 2\% during the period to 18\%, the language mismatch remains to be at the same ratio. Clearly our post-training methods are not designed to handle these specific failure categories. In contrast, the EBR usage in the group search product is seeing an upward trending from 60\% to 70\%. The reason for the change is diversified, like inventory shift and text based retrieval being removed to save infra cost, etc. We still need to put focus on the problem and if possible, more targeted solution on each of the failure categories in the long run.

\section{conclusions and future work}
In this paper we explicitly call out two failure categories introduced by applying embedding-based retrieval, and several techniques to address the issue with offline and online evaluation. There are several potential directions to explore. The user cohort modeling in our linear regression model is preliminary. Dimensions like user demo-graphical feature, social context feature, preference and search history are not well studied. Feature selection and regularization are not included. A more direct approach is to understand how deep retrieval model behave in different user cohorts and improve directly in the model training stage, for these different failure categories.

\section*{Acknowledgments} 
The author would like to thank Haocheng Wu and Qing Liu for their generous help during the development, discussion and experimenting of this paper.

\section*{COMPANY PORTRAIT}
Meta Platforms, Inc. is an American technology company based in Menlo Park, California. The company owns Facebook, one of the largest social media and social networking service, among other products and services. It strives to give people the power to build community, bring the world closer together.

\section*{PRESENTER’S BIO}
Wenping Wang is a software engineer at Facebook. His team is responsible for Facebook search's quality, the retrieval, ranking, etc. Before that, he received Master's degree from Carnegie Mellon University. He is interested in computer systems, machine learning in general and their real-world applications.

\bibliographystyle{ACM-Reference-Format}
\bibliography{references}

\end{document}